\documentstyle[12pt]{ioplppt}

\newcommand{\CS}{{\em Complex Systems} }
\newcommand{\DAM}{{\em Discr.~Appl.~ Math.} }
\newcommand{\JTB}{{\em J.~Theoret.~Biol.} }
\newcommand \real   {I\kern-1.5mm{R}}
\newcommand \nat    {I\kern-1.5mm{N}}
\newcommand \sign   {\mathop{\rm sign}\nolimits}

\newcommand \beq {\begin{equation}}
\newcommand \beqar {\begin{eqnarray}}
\newcommand \eeq {\end{equation}}
\newcommand \eeqar {\end{eqnarray}}
\newcommand \refeq [1]{\mbox{(\ref{#1})}}

\def \R{{\rm I\kern -0.17em R}}
\def \N{{\rm I\kern -0.17em N}}
\def \Q{{\rm\kern 0.26em \vrule height 1.6ex depth -0.1ex
 width 0.05em \kern -0.31em Q}}
\def \C{{\rm\kern 0.26em \vrule height 1.5ex depth -0.1ex
 width 0.05em \kern -0.31em C}}
\begin{document}
\jl{1}
\letter{Limit cycles of a perceptron}
\author{M. Schr\"oder and W. Kinzel}
\address{Institut f\"ur Theoretische Physik, Universit\"at 
W\"urzburg,
Am Hubland, D-97074 W\"urzburg}

\begin{abstract}
An artificial neural network can be used to generate
a series of numbers. A boolean perceptron generates bit
sequences with a periodic structure.
The corresponding spectrum of cycle lengths is investigated
analytically and numerically; it has similarities with
properties of rational numbers.
\end{abstract}

In the last fifteen years models and methods of statistical physics have
successfully been used to understand emergent computation of neural
networks. Several properties of infinitely large attractor and multilayer
networks could be calculated analytically. Such systems of simple units
interacting by synaptic weights can be used as associative memory and
classifiers; they are trained by a set of examples, detect unknown
rules and structures in high--dimensional data, and store patterns in a
distributed and content addressable way (Hertz \etal 1991, Watkin \etal 1993,
Opper and Kinzel 1996).

Another important application of neural networks is time series analysis
(Weigand 1993). But only recently statistical physics has been used to model
training and prediction of bit sequences by a perceptron (Eisenstein \etal 1995,
Schr\"oder \etal 1996). A neural network is trained by a sequence of numbers; 
after the training phase the network makes predictions on the rest of the
sequence. In analogy to generalization the training and test data are
generated by a neural network, as well.  It turns out that the generation
of sequences of numbers by a perceptron or multilayer network is already
an interesting problem which should be understood before prediction is
investigated (Eisenstein \etal 1995, Kanter \etal 1995). 

This problem is a special case of 
the {\sl neuronic equations} of Caianiello (1961). 
These equations, which were suggested to model a neuron including time 
dependency, can only be solved in special cases.
Most work was done for one input and a memory back into time with
couplings that decrease exponentially $w_i=a^i$ ($a<1$). 
Several analytic results for the transients and the limit cycles of
the resulting dynamics have been achieved for this case (for example
Caianiello and Luca 1965 or Cosnard \etal 1988). 
Only few work has been done for other
weights (for example Cosnard \etal 1988b and 1992).
Our motivation for studying this recursion equation is to examine
the generalization ability and in a first step the ability of a
perceptron to generate time series. 
Hence we have no restrictions on the weight vector a priori.

Numerical analysis of a perceptron with random weights 
generating sequences of numbers shows
that the sequences are related to the Fourier modes of the weight vector.
Therefore it is useful to study weight vectors with a single mode only. In
this case an analytic solution of a stationary sequence could be derived
for large frequencies (Kanter \etal 1995). 

This solution holds for continuous odd transfer functions, for example
$\tanh (\beta x)$. As a function of the slope $\beta$ a phase transition 
to a nonzero
sequence occurs. The phase of the weight vector results in a frequency
shift of the attractor. In this letter we want to extend this solution to
infinite slope $\beta$ and general frequencies, that is we derive an 
analytic solution for the bit
generator. We find a much richer structure of the bit sequences generated
by a boolean  perceptron compared to sequences of continuous ones. 

A bit generator (figure 1) is defined by the equation 

\begin{equation} 
S_\nu = \sign \,
\sum\limits^{N}_{j=1} \, w_j S_{\nu-j} \qquad (\nu\in\{N, N+1, N+2, \dots\})
\label{bgint}
\end{equation} 

where the $S_\nu \in \{+1, -1\} $ is a bit of the
sequence $(S_0, S_1, S_2, S_3,\dots)$ and $\underline{w} \in \real^{N}$ is
the weight vector of the perceptron of size $N$. As mentioned, it is
useful to restrict the weights to one single mode.

\begin{equation} 
w_j = \cos (2\pi q \, \frac{j}{N} + \pi \phi )
\end{equation} 

where $q \in
\nat$ is the frequency and $\phi \in [0,1[$ is the phase of the weights.
Given an initial state $(S_0, \dots, S_{N-1})$ equation (1) defines a sequence
$(S_N, S_{N+1}, \dots)$ which has to run into a periodic cycle of length $L
\le 2^N$. We try to find an analytic solution of the periodic attractor. 

Equation (1) may be expressed in terms of the local fields $h_\nu = {1\over N}
\sum_{j=1}^N w_j S_{\nu-j}$: 
\begin{equation} 
h_\nu = \frac{1}{N} \sum\limits^{N}_{j=1} \, \cos
\left( 2\pi q \, \frac{j}{N} + \pi \phi \right) \sign (h_{\nu-j}).
\label{bg1}
\end{equation} 
We have to solve this self-consistent equation for the function $h_\nu$.
Since simulations show that limit cycles are dominated by one frequency,
we assume $\sign(h_\nu) $ is a periodically alternating step function with
frequency $k+\tau$ (with $k \in \nat, \tau \in [0,1[$), where
the frequency is defined for the variable $\nu/N$:
\beq
\sign(h_\nu)=\sign(\sin(2\pi {(k+\tau)\over N} \nu)).\label{sinus}
\eeq
In the case that $N$ is a multiple of $2(k+\tau)$, i.e.\ for integer
wavelengths, we
found an analytic solution of equation (1).
With this ansatz the right hand side of equation \refeq {bg1}
is a periodic function with a period of length $N/(k+\tau)$ and
$h_\nu=-h_{\nu+N/(2(k+\tau))}$ for $\nu\in\{0,\dots,{N\over2(k+\tau)}-1\}$.
Our main result is

\beqar
\fl h_\nu={1\over N\sin({\pi q\over N})}
{\sin({1\over2}(T+1)({\pi q\over k+\tau}+\pi))
\over \cos({\pi q\over2(k+\tau)})}\cdot
\sin(2\pi q{\nu\over N}+\phi\pi+{T\over2}({\pi q\over k+\tau}+\pi)+
{\pi q\over N})
\nonumber\\
\lo -\cases{0&for $T$ odd\cr
+{2\over N}\cos(\phi\pi)+
{\sin(\phi\pi-{\pi q\over N})\over N\sin({\pi q\over N})}&for $T$ even\cr}.
\label{fi}
\eeqar
where we have abbreviated $T=[(k+\tau)(2-2\nu/N)]$ using the 
{\sl Gaussian bracket} $[x]$ that denotes the closest integer less than $x$.
For $\phi=\tau=0$ and $k=q$ this results in
\beq
h_\nu={2k\over N \sin({\pi k\over N})}\sin({2\pi k\nu\over N}+
{\pi k\over N})
\eeq
A sample function is plotted in figure 2. 
Note that it consists of two parts with the same frequency $q$.
In the limit $N\to\infty$ these parts are connected continuously.
Equation \refeq{bg1} gives the condition 
\beq
h_\nu\ge0 \qquad \nu\in\{0,\dots,{N\over2(k+\tau)}-1\}\label{sig}
\eeq
Figure 3 shows the possible frequencies that satisfy condition \refeq{sig} 
within our ansatz. Only values of $k+\tau=N/(2i)$ with integers $i$ are
possible with our ansatz.

For $N\to\infty$ a necessary condition for equation \refeq{sig} is $h_0=0$.
For $k\ge q$ this is sufficient so the nontrivial 
$(h_\nu\not\equiv0)$
solutions are given by the values of $\tau,k$ that fulfill the equation
$ \sin(\phi\pi+(2k+1)({\pi q/(2(k+\tau))}+{\pi/2}))=0$
which is equivalent to $q(2k+1)=(k+\tau)(2z-2\phi-1)$ with an integer $z$.
The frequencies $k+\tau$ that are allowed from this condition
are shown in figure 3, too.

We see that the analytic solution of the sequence generator with a
continuous transfer function (Kanter \etal 1995) cannot just be 
extrapolated to the case of the bit generator. 
The continuous generator, close to the
transition point and for $k\gg1$, has $k=q$ and $\tau=\phi$, 
whereas we find a spectrum of solutions with $k \ge q$ and 
$\tau (q,\phi,k)$, as shown in figure 3. 

Up to now we have considered only integer wavelengths. Now we
want to discuss the general case of arbitrary values of $q$ and $\phi$.
We want to address the two questions:

\begin{enumerate}
\item Do additional solutions consisting of one frequency exist?
\item What are the properties of the bit sequences? 

\end{enumerate}

We assume that there are solutions of the form \refeq{sinus} with
general frequencies $k+\tau$ for a given system size $N$.
As a function of $q+\phi$ we numerically scan the output frequency $k+\tau$ 
and determine the frequency of the limit cycle when the system was
started with a sequence of frequency $k+\tau$.
Some of these initial states stay at stable states with almost the
same frequency.  Other ones run
to the lowest branch ($k=q$). Random initial states lead to the lowest branch
with a very hight probability.
Figure 3 shows that the results of this simulation are in 
agreement with the extension of our equations \refeq{fi} and \refeq{sig} 
to general $k+\tau$ which
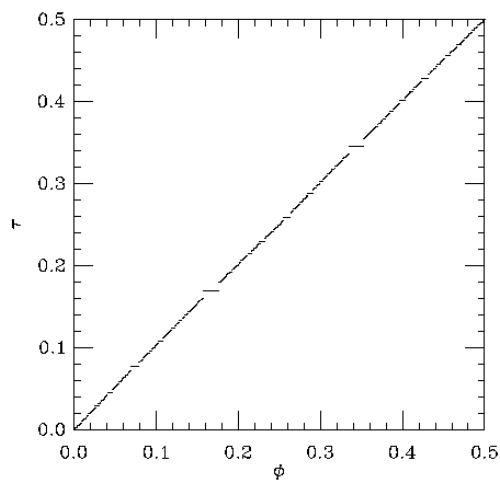
leads to allowed regions for for $q+\phi$ as a function of $k+\tau$. 
For the lowest branch $k=q$ the phase $\phi$ of the weights
results in a frequency shift $\tau$ of the bit sequence, 
with $\tau\approx\phi$ for $q\gg1$ similar to the continuous case.

The next problem is to understand the length $L$ of the stationary cycle
generated by the finite bit generator 
with one frequency in the couplings and a random initial vector. 
We consider the case $q=k$, only.
Figure 5 shows the results of a
numerical calculation of Equation (1) for $N =1024$. Obviously $L$ has a rich
structure as the function of the phase $\phi$ of the weights. For $0 \le
\phi \le \frac{1}{2}$ each cycle has only one maximum in the Fourier
spectrum with frequency $q + \tau (\phi)$. The numerical results show that
in this case $L$ is limited to the value $2N$.  Each value of $L$ belongs
to a whole interval in the $\phi$--axis, but only to a single value of
$\tau$. Hence, the function $\tau (\phi)$ has a step like structure;
$\tau$ is locked at rational numbers as shown in Figure 6. The size of
the steps decreases with increasing $N$ and $r$.

Can this structure of $L(\tau)$ be understood from the extension of
the analytic solution Equation \refeq{fi}? For arbitrary values of $\tau$ the
sequence $S_l$ is quasi-periodic, in general with an infinite period $L$.
However, if $k+ \tau$ is rational, $k + \tau = r/s$ with integers $r$ and
$s$ which are relatively prime, then the period $L$ is given by the numerator
of $Ns/r$. 
This means, that $L$ is the smallest multiple of the wavelength $N/(k+\tau)$
that is an integer.
In fact, in Figure 5 we have plotted all of these values of $L$
for $q+\phi = r/s$ and $L < 2N$. For $L<N$ all of these $L$ values
correspond to a cycle of the bit generator. For $N< L < 2N$ the bit
generator produces only even values of $L$ whereas the analytic argument
gives all integers $L$. 

For $\frac{1}{2} < \phi <1$ the bit-generator essentially produces the same
structure of $L$ values as for $0<\phi<\frac{1}{2}$. However, there are
always a few solutions which are mixtures of several modes $k_1 + \tau_1$
and $k_2 + \tau_2$, and which yield periods with $L >2N$. For $0<\tau<
\frac{1}{2}$ we never observed such mixtures of modes. 

As a side remark we notice that the structure of $L(q+\tau)$ is
essentially determined by properties of rational numbers, which might be
discussed in high school mathematics. If the numerator $p$ is plotted for
each rational number $x = p/r$ in the unit interval, we obtain Figure 7
($r<800$). Hence, above each rational number
$p/r$ a roof opens below which no other values of $p$ appear. Each roof
has the form $1/|p-rx|$. Low values of $r$ have a wide roof.
These results, which determine the structure of the cycle lengths of the
bit generator, may be well known in number theory and nonlinear dynamics
(circle map, winding number), but they have skipped our attention so far. 

The upper bound of the cycle length $L$ of $2N$ can be understood as follows.
The analytic solution \refeq{fi}, extended to general values of 
$k+\tau$, yields quasi periodic bit sequences with infinite cycle lengths 
$L = \infty$ for irrational $k+\tau$. 
However, each cycle length has to be limited by the number of
input strings for the deterministic bit generator, Equation \refeq{bgint}, 
which gives 
$L < 2^{N}$. The last argument opens a different possibility to calculate $L$
from equation \refeq{fi}. 
Let us start with the sequence $(S_0, S_1, \dots, S_{N-1})$
given by the equation \refeq{fi}, of the analytic solution.  
If a bit generator tries to follow this solution 
it can do so only if
each input string $(S_l, S_{l+1}, \dots, S_{l+N-1})$ has not occurred before.
Hence, the first appearance of a previous sequence

\beq
(S_l,S_{l+1},\dots,S_{l+N})=(S_{l+L},S_{l+L+1},\dots,S_{l+L+N-1})
\eeq
defines a length $L$ of a cycle. 

More insight can be achieved by examining the continued fraction expansion
of $2(k+\tau)/N$:

\beq
2{k+\tau\over N}={1\over a_1+{1\over a_2+{1\over a_3+\dots}}},\qquad a_i\in\N
\eeq

We define the expansion to order $i$ to be $s_i$ and
$l_i$ to be the denominators the $s_i$:
\beq
\begin{array}{rclrcl}
s_0&=&0&\qquad l_0&=&1\cr
s_1&=&{1\over a_1}&\qquad l_1&=&a_1\cr
s_2&=&{1\over a_1+{1\over a_2}}&\qquad l_2&=&a_1a_2+1\cr
s_3&=&{1\over a_1+{1\over a_2+{1\over a_3}}}&\qquad l_3&=&(a_1a_2+1)a_3+a_1\cr
&\vdots&&\vdots&\cr
&&&l_i&=&a_il_{i-1}+l_{i-2}
\end{array}
\eeq

If $2(k+\tau)/N$ is of the form $s_1=1/a_1$ the length $L$ of the 
cycles as a function of $N$ can be given easily:
\beq
L=\cases{1&for $N<a_1$\cr2a_1&for $N\ge a_1$\cr}
\eeq
Is is obviously limited by 2N.

For general $2(k+\tau)/N$ the continued fraction expansion reveals a 
hierarchy of of ``defects'', each having a period of $l_i$, in the 
periodic structure of the resulting sequence.
This leads to the the fact that for any length scale introduced
by $N$, two identical subsequences of length $N$ with a distance less than $2N$ 
can be found.

As a consequence the perceptron locks, for a given frequency $k+\tau$,
in cycles that correspond
to frequencies given by the continued fraction expansion of $2(k+\tau)/N$, 
truncated at a certain depth.
This explains the steps in figure 6.

Finally we point out similarities to the case of bit generators with
exponentially decaying weights and additional bias (Cosnard \etal 1988).
In this case one finds cycles which are limited by $N+1$. All of the
cycles can be classified by rational numbers $r/L$ where $L$ is the
length of the cycle, $L\le N+1$ and $r$ is the number of positive bits in the
cycle.

In summary, we have obtained an analytic solution 
for the cycles of a bit-generator with periodic weight vectors.
We found a whole spectrum of periodic
attractors; the frequencies $k + \tau$ depend in a complex way on the
frequency $q$ and phase $\phi$ of the weight vector of the perceptron. 

Numerical simulations showed that the bit sequences relax
into cycles with lengths $L$, which are smaller than $2N$. The structure
of $L$ as a function of $k+\tau$ has been analyzed in terms of number
theory. An analytic solution was given for certain frequencies; the
extension to the measured frequencies results in
a similar structure of cycles. 

\section*{Acknowledgments}
This work has been supported by the Deutsche Forschungsgemeinschaft.
We thank Ido Kanter for valuable discussions.
We also thank a referee for pointing out the interesting papers of 
Cosnard \etal to us which are related to our work.

\References
\item[] Caianiello E R 1961 \JTB {\bf 2} 204
\item[] Caianiello E R and De Luca A 1965 {\sl Kybernetik} {\bf 3} 33
\item[] Cosnard M, Tchuente M and Tindo G 1992 \CS {\bf 6} 13
\item[] Cosnard M, Goles Chacc E and Moumida D 1988 \DAM {\bf 21} 21
\item[] Cosnard M, Moumida D, Goles E and St.Pierre T 1988 \CS {\bf 2} 161
\item[] Eisenstein E, Kanter I, Kessler D A and Kinzel W 1995 \PRL {\bf 74} 6
\item[] Hertz J, Krogh A and Palmer R G 1991 {\it Introduction to
the theory of neural computation} (Redwood City, CA: Addison Wesley)
\item[] Kanter I, Kessler D A, Priel A and Eisenstein E 1995 \PRL {\bf 75} 
2614
\item[] Opper M and Kinzel W 1996 {\it Statistical Mechanics of Generalization}
in Domany E, van Hemmen J L and Schulten K (eds) {\it Physics of Neural 
Networks III} (New York: Springer)
\item[] Schr\"oder M, Kinzel W and Kanter I 1996 \JPA {\bf 29} 7965
\item[] Watkin T L H, Rau A and Biehl M 1993 \RMP {\bf 65}(2) 499
\item[] Weigand A S and Gershenfeld N A (eds) 1993 {\it Time Series 
Prediction, Forecasting the Future and Understanding the Past} 
(Santa Fe: Santa Fe Institute) 
\endrefs
\end{document}